\documentclass[11pt]{article}

\usepackage{pb-diagram}
\usepackage{latexsym}
\usepackage{amsfonts}
\usepackage[usenames,dvipsnames]{xcolor}
\usepackage{graphicx,amssymb,amsmath,epsfig}
\usepackage[english]{babel}
\usepackage{graphicx}% Include figure files
\usepackage{dcolumn}% Align table columns on decimal point
\usepackage{bm}

\definecolor{myblue}{rgb}{0,0,0.8}
\usepackage[colorlinks=true
,urlcolor=myblue	
,anchorcolor=myblue
,citecolor=myblue
,filecolor=myblue
,linkcolor=myblue
,menucolor=myblue
,pagecolor=myblue
,linktocpage=true  %%%%%%  Link no numero da pagina em vez do titulo   %%%%%%%%%%
]{hyperref} 

\definecolor{green}{RGB}{0, 130, 0}
\definecolor{grey}{RGB}{90, 90, 90}
\catcode`\@=11
\def\marginnote#1{}

\newcount\hour
\newcount\minute
\newtoks\amorpm
\hour=\time\divide\hour by60 \minute=\time{\multiply\hour by60
\global\advance\minute by-\hour}\edef\standardtime{{\ifnum\hour<12
\global\amorpm={am}%
        \else\global\amorpm={pm}\advance\hour by-12 \fi
        \ifnum\hour=0 \hour=12 \fi
        \number\hour:\ifnum\minute<10
0\fi\number\minute\the\amorpm}}
\edef\militarytime{\number\hour:\ifnum\minute<10 0\fi\number\minute}

\def\draftlabel#1{{\@bsphack\if@filesw {\let\thepage\relax
   \xdef\@gtempa{\write\@auxout{\string
      \newlabel{#1}{{\@currentlabel}{\thepage}}}}}\@gtempa
   \if@nobreak \ifvmode\nobreak\fi\fi\fi\@esphack}
        \gdef\@eqnlabel{#1}}
\def\@eqnlabel{}
\def\@vacuum{}
\def\draftmarginnote#1{\marginpar{\raggedright\scriptsize\tt#1}}
\def\draft{\oddsidemargin -.5truein
        \def\@oddfoot{\sl preliminary draft \hfil
        \rm\thepage\hfil\sl\today\quad\militarytime}
        \let\@evenfoot\@oddfoot \overfullrule 3pt
        \let\label=\draftlabel
        \let\marginnote=\draftmarginnote

\def\@eqnnum{(\theequation)\rlap{\kern\marginparsep\tt\@eqnlabel}%
\global\let\@eqnlabel\@vacuum}  }

\def\numberbysection{\@addtoreset{equation}{section}
        \def\theequation{\thesection.\arabic{equation}}}

\def\underline#1{\relax\ifmmode\@@underline#1\else
 $\@@underline{\hbox{#1}}$\relax\fi}

\catcode`@=12 \relax

\numberbysection
\topmargin by -\headsep
\def\nonu{\nonumber}
\def\br{\begin{eqnarray}}
\def\er{\end{eqnarray}}
\def\lb{\lbrack}
\def\rb{\rbrack}

\def\({\left(}
\def\){\right)}
\def\[{\left[}
\def\]{\right]}

\def\lie{{\cal G}}
\def\a{\alpha}
\def\b{\beta}

\def\d{\delta}

\def\eps{\epsilon}

\def\l{\lambda}
\def\L{\Lambda}

\def\pa{\partial}

\def\s{\sigma}

\def\tp0{\Theta_{+}^{(0)}}
\def\tm0{\Theta_{-}^{(0)}}

\def\cm{{\cal M}}

\def\l{\lambda}

\def\nonu{\nonumber}

\def\bi{\begin{itemize}}
\def\ei{\end{itemize}}

\def\e{\epsilon}
\def\ck{{\cal K}}
\newcommand{\grad}[1]{{}^{(#1)}}

\begin{document}

%\today 
%\hskip 5cm generalized-B\"acklund-JPA-revised-30-12.tex
\vspace*{1cm}
\noindent

\vskip 1 cm
%%%%%%%%%%%%%%%%%%%%%%%%%%%%%
%%%%%%%%%%%%%%%%%%%%%%%%%%%%%%%%%%%%%

%\vskip 1 cm
\begin{center}
{\Large\bf   Generalized   B\"acklund transformations for Affine Toda Hierarchies}
\end{center}
\normalsize
\vskip 1cm
\begin{center}
{J.M. de Carvalho Ferreira}\footnote{\href{mailto:jogean.carvalho@unesp.br}{jogean.carvalho@unesp.br}},   J.F. Gomes\footnote{\href{mailto:francisco.gomes@unesp.br}{francisco.gomes@unesp.br}}, G.V. Lobo\footnote{\href{mailto:gabriel.lobo@unesp.br}{gabriel.lobo@unesp.br}}, and A.H. Zimerman\footnote{\href{mailto:a.zimerman@unesp.br}{a.zimerman@unesp.br}}\\[.7cm]

\par \vskip .1in \noindent
 {Instituto de F\'isica Te\'orica - IFT/UNESP,\\
Rua Dr. Bento Teobaldo Ferraz, 271, Bloco II,
CEP 01140-070,\\ S\~ao Paulo - SP, Brasil.}\\[0.3cm]

\vskip 2cm

\end{center}

\begin{abstract}

    The construction of generalized B\"acklund transformation for the  $A_n$ Affine Toda hierarchy is proposed in terms of gauge transformation acting on the zero curvature representation.  Such construction  is based upon the graded structure of the underlying affine algebra which induces a classification of generalized B\"acklund transformations.  Moreover, explicit examples for { $sl(3)$ and $sl(4)$} lead to uncover interesting composition properties of various types of B\"acklund transformations. The universality character of  the gauge-B\"acklund transformation method is extended to all equations of the  hierarchy.  Such interesting property provides  a systematic framework to construct B\"acklund transformations to higher flow equations. Explicit example for the  simplest higher  flow  of the $sl(3)$ hierarchy is presented.

\end{abstract}

\section{Introduction}

Integrable hierarchies  provide  interesting prototypes  where   an underlining  affine algebraic structure  plays a crucial ingredient either   in constructing    and classifying time evolution equations in terms of Lax pairs  $A_{\mu} (\varphi_i) \in  \hat \lie $, $\mu =0,1$ (see e.g.\cite{jfg-rev} for review) or in the construction of  soliton (multi) solutions  from tau functions  and representation theory   of such   algebras \cite{babelon}. Examples  of well known (relativistic) integrable models as the sine (sinh)-Gordon, Tzitz\'eica, Lund-Regge  and other non relativistic integrable models as generalized Non-Linear Schr\"odinger (NLS), mKdV, etc have been extensively studied within such context (see for instance  \cite{lr}-\cite{assis}).

On the other hand, the existence of B\"acklund transformations   is  a peculiar feature of integrable models which relate two field configurations of a nonlinear integrable equation. Originally this was  employed to    construct (multi) soliton solutions  from a nonlinear superposition principle (see \cite{rogers} for a review). More recently,   B\"acklund transformations (BT) have been employed to describe  integrable defects \cite{corr1-2004}-\cite{ymai} in the sense that  two field configurations are connected  by  a defect. This idea was first proposed in \cite{corr1-2004} for the Liouville and sine-Gordon theories.  {In such cases, the integrability is preserved if the two solutions   at the defect are connected by the B\"acklund transformation.}

In this paper  we  discuss a systematic construction  of  the various types of B\"acklund transformation for   $  \lie =A_n$ affine Toda hierarchy.  We   construct a B\"acklund-gauge transformation acting  on the Lax pair preserving the zero curvature representation and  hence the equations of motion. This approach have been  applied to the mKdV hierarchy associated to the affine algebra $\hat {sl}(2)$ in \cite{ana1} and  \cite{ana-miura} where  the B\"acklund-gauge transformation was  shown  to be universal for  all equations  within  the hierarchy. Type I BT was first constructed for the sine-Gordon  model more than  a hundread years ago  and was generalized  for the  $  \lie =A_n$ Toda theories by Fordy and Gibbons in ref. \cite{fordy}. Moreover a new type of B\"acklund transformation, dubbed, type II, involving an auxiliary field  was proposed  in \cite{corrigan}  for the Tzitz\'eica model. The object of this paper is to classify and construct the various  types of B\"acklund transformations in terms of an  affine  graded structure. More recently the generalization of BT  have been studied in connection to symmetries (folding of Dynkin diagrams) \cite{robertson} and to other  peculiar properties of the underlying algebraic structure (other than $A_n$)  \cite{bristow}.

In section 2  we  define the model in terms of  zero curvature  representation
\br
\pa_x A_t (\varphi_i) -\pa_{t} A_x(\varphi_i) + [A_x(\varphi_i), A_t(\varphi_i)] =0,   \label{01}
\er
                by constructing the two dimensional gauge potentials    $A_{\mu} \in \hat \lie$ ( {we denote $A_{\mu =0 }\equiv A_t$ and $A_{\mu =1} \equiv A_x$ }).  We then   construct the type I   and type II BT  connecting two sets of solutions, $\{\phi_i\}$ and $\{\psi_i\}, \quad i=1,\cdots,n$ of the same eqns. of motion for the affine  Lie algebra $ \hat \lie =\hat {sl}(n+1)$ in terms  of B\"acklund-gauge Transformation (BT). Such transformation is generated by a group  element $U(\phi_i, \psi_i, \l)$   acting  on the gauge potentials (Lax pair)  $A_{\mu} (\varphi_i ) $, i.e.,
\br
U(\phi_i, \psi_i,\l)A_{\mu}(\phi_i)= A_{\mu }(\psi_i) U(\phi_i, \psi_i,\l) + \pa_{\mu} U(\phi_i, \psi_i,\l), \qquad \mu = 0,1.
\label{004}
\er
The various BT  are  dependent upon a parameter $\l$  and  are  shown to be classified according to  the principal gradation  explained and developed in the Appendix A.

Next in section 3 we discuss explicitly the construction of type II BT for $sl(3)$  model and  show that the natural choice of  parameters  induces    $U(\phi_i, \psi_i, \l)$  to be of unit determinant. Moreover, eqn. (\ref{004})  shows that both, $U(\phi_i, \psi_i,\l)$ and $\tilde U(\phi_i, \psi_i,\l) \equiv    U^{-1}(\psi_i, \phi_i,\l)$  satisfy our requirements  (\ref{004}). In particular    for the type II BT we verify that, 
\br
U(\phi_i, \psi_i,\l) = c \, U^{-1}(\psi_i, \phi_i,\l), \quad c = {\rm const} \label{inverse}
\er  
for $\lie =sl(3)$. 
Such condition indicates  that, for this case,  the type II B\"acklund-gauge  transformation can  be entirely parametrized  in terms of 2  auxiliary fields,  $\L_i, i=1,2$.    The Tzitz\'eica limit is then taken  and  shown to reproduce  the B\"acklund equations  proposed  in \cite{corrigan}. Moreover we discuss the  composition  structure of these  BT and show how these  different types can be interrelated.

In section 4 we propose the type II  B\"acklund-gauge transformation  for the general case $\lie = sl(n+1)$  in terms of $n $   auxiliary fields, $\L_i, \; i=1, \cdots, n$  and verify its inverse relation (\ref{inverse}) for $n=2$ and $3$.  We  then extend  the argument to other nonlinear eqns. of the same hierarchy (higher flows). As an example we construct explicitly the first positive  grade  time evolution equation in section 5. Examples of solutions  of vaccum-one soliton and scattering of one soliton solutions are  discussed in section 6. Finally in the Appendix A we discuss the $\hat sl(n+1)$ decomposition according to the principal gradation.

\section{The Model}

Consider the Lax pair given by
\br
A_x = E^{(1)} + B^{-1}\pa_x B , \qquad A_{t} = B ^{-1}E^{(-1)} B.
\label{1}
\er
The Zero Curvature Condition (ZCC)  leads to the  Affine Toda  (AT)  eqns. of motion, 
\br
\pa_x A_t -\pa_{t} A_x + [A_x, A_t] =- \pa_t  (B^{-1}\pa_x  B ) + [E^{(1)},  B^{-1}E^{(-1)} B] =0
\label{2}
\er
 {For the $sl(n+1)$ case the  fields of the theory, $\phi_i,  n=1, \cdots n$ parametrize a zero grade  group element $ B= e^{\sum_{i=1}^{n}\phi_i h_i}$, where $h_i$  are the Cartan subalgebra generators  and $E^{(\pm 1)} = \sum_{i=1}^{n} E_{\pm \a_i} + \l^{(\pm)}E_{\mp (\a_1 + \cdots + \a_n)}$,   $\a_i, i=1,\cdots,n$ are the simple roots of $sl(n+1)$ and $\psi = \a_1 +\a_2 + \cdots + \a_n$ is the highest root (see Appendix A ). }In components eqn. (\ref{2}) yields the affine Toda equations,
\br
\pa_x\pa_t \phi_i =e^{K_{ij}\phi_j} - e^{-(K_{1j} + K_{2j} + \cdots + K_{nj})   \phi_j } \quad i=1,\cdots,n
\label{3}
\er
where $K  $ denotes the Cartan Matrix of $sl(n+1)$, $K_{ij} = {{{2\a_i\cdot \a_j}\over {\a_j^2}}} \quad i,j=1,\cdots, n$.\\
In this paper we shall discuss  the construction of B\"acklund-gauge Transformations (BT), $U(\phi_i, \psi_i)$  acting  on the Lax pair (\ref{1}) satisfying (\ref{004}) interpolating    between two  field configurations, $\{ \phi_i\} $ and $\{ \psi_i \}    $   of the same equation of motion \footnote{Here $x$ and $t$ denote the light cone coordinates}.

We now analyse the various  possibilities  leading to B\"acklund transformations (BT) which will be classified according  to its graded structure  given in Appendix A.

\subsection{Type I B\"acklund Transformation} 

Let us consider the first case  in which  $U$ can be decomposed  as  sum of grades zero and  $-1$ components as $U ={U\grad{0} + \sigma U\grad{-1}}$. According to the grading structure given in the appendix A, 
\begin{eqnarray}
U_{a,b} = {u\grad{0}_a} \, \d_{a,b}+ {\sigma u\grad{-1}_{b}}\,\d_{a,b+1} + \lambda^{-1}\,{\sigma u\grad{-1}_{n+1}}\,\d_{a,1}\d_{b,n+1}, \quad a,b=1, \cdots ,n+1
\label{matrixu}
\end{eqnarray}
or in matrix form 
\begin{eqnarray}
U (\phi_i,\psi_i)= 
\left( {\begin{array}{cccc}
	{u\grad{0}_1} & 0 & \cdots & 0 \\
	0 & {u\grad{0}_2} & \cdots & 0 \\
	\vdots & \vdots & \ddots   & \vdots  \\
	0&0 &\cdots & {u\grad{0}_{n+1}}
	\end{array} } \right) + 
\left( {\begin{array}{cccc}
	0 & \cdots & 0 & \l^{-1}{\sigma u\grad{-1}_{n+1}} \\
	{\sigma u\grad{-1}_1} & \cdots & 0 & 0 \\
	\vdots & \ddots &  \vdots  & \vdots  \\
	0& \cdots&{\sigma u\grad{-1}_n} & 0
	\end{array} } \right).
\label{6}
\end{eqnarray}
From (\ref{1}) consider  the  components   of the Lax pair,
\begin{equation*}
\begin{split} 
A_x (\phi)_{a,b} &= \pa_x ( \phi_a-\phi_{a-1}) \d_{a,b}+ \d_{a,b-1} + \l  \d_{a,n+1} \d_{b,1}\\[8pt]
%\label{ax}
%
A_{t}(\phi )_{a,b} &= e^{-\phi_{b-1}+2\phi_b-\phi_{b+1}}\d_{a-1,b} + \l^{-1} e^{-\phi_1-\phi_n}\d_{a,1}\d_{b,n+1}, 
%\label{at}
\end{split}
\end{equation*}
where $a,b=1,\cdots, n+1$ and $\phi_0= \phi_{n+1}=0$, which in matrix form can be written  as
\begin{equation*}
A_x(\phi) =  \pa_x \phi_i h_i + E^{(1)} =
\left( {\begin{array}{ccccc}
	\pa_x \phi_1 & 1 &  0 & \cdots & 0 \\
	0 & \pa_x(\phi_2-\phi_1) & 1 &  &  \\
	&&&&\vdots\\
	0 & 0 &  & \cdots &  \\
	\vdots & \vdots &  & \pa_x (\phi_n-\phi_{n-1}) &   1  \\
	\l & \ldots &  & 0 & -\pa_x \phi_n
	\end{array} } \right), 
\label{7}
\end{equation*}
and 
\begin{equation*}
A_{t}(\phi) = B^{-1} E^{(-1)} B = \left( {\begin{array}{ccccc}
	0 & 0 & \cdots & 0& \l^{-1}e^{-\phi_1 - \phi_n} \\
	e^{2\phi_1-\phi_2} & 0 & \cdots & 0 &   0 \\
	0 & e^{-\phi_1+2\phi_2- \phi_3} &	& \vdots &     \vdots  \\
	\vdots & \vdots & \ddots & 0 & 0\\
	0& 0 & & e^{-\phi_{n-1}+2\phi_{n}} & 0
	\end{array} } \right).
\label{at}
\end{equation*}

For $A_{\mu} = A_x$,   eqn. (\ref{004})  rewritten in the  matrix form $[U A_x(\phi)-A_x(\psi)U-\pa_xU]_{ik}=0$ leads to  the following eqns.,
\begin{equation}
\begin{split}
 \(  -\pa_x {u\grad{0}_i}+{u\grad{0}_i}\pa_x(\phi_i-\phi_{i-1}-\psi_{i}+\psi_{i-1})+{\sigma}\({u\grad{-1}_{i-1}} - {u\grad{-1}_i}\) \) \d_{i,k}& \\
+{\sigma}\( -\pa_x {u\grad{-1}_{i-1}} +{u\grad{-1}_{i-1}}\pa_x(\phi_{i-1}-\phi_{i-2}-\psi_{i}+\psi_{i-1} )\) \d_{i,k+1}&\\
-\lambda^{-1}{\sigma}\(\pa_x{u\grad{-1}_{n+1}}+{u\grad{-1}_{n+1}}\pa_x(\phi_n+\psi_1)\)\d_{i,1}\d_{k,n+1}&  \\
+({u\grad{0}_i}-{u\grad{0}_{i+1}})\d_{i,k-1} +\l({u\grad{0}_i}-{u\grad{0}_k})\d_{k,1}\d_{i,n+1} &=0 \\
\end{split}
\label{x01}
\end{equation}
for $i,k=1, \cdots, n+1$, and ${u\grad{-1}_{0}}= {u\grad{-1}_{n+1}}$. Similarly  for $A_{\mu} = A_t$, $[U A_t(\phi)-A_t(\psi)U-\pa_tU]_{ik}=0$  yields,
\begin{equation}
\begin{split}
\(-{\sigma}\pa_t {u\grad{-1}_{i-1}}+{u\grad{0}_i}e^{-\phi_{i-2}+2\phi_{i-1}-\phi_{i}} - {u\grad{0}_{i-1}}e^{-\psi_{i-2}+2\psi_{i-1}-\psi_{i}}\)\d_{i,k+1} & \\
+{\lambda^{-1}}\(-{\sigma}\pa_t {u\grad{-1}_{n+1}}+{u\grad{0}_{1}}e^{-\phi_1-\phi_n} -{u\grad{0}_{n+1}}e^{-\psi_1-\psi_n}\)\d_{i,1}\d_{k,n+1} &\\
+{\sigma}\({u\grad{-1}_{i-1}}e^{-\phi_{i-3}+2\phi_{i-2}-\phi_{i-1}}-{u\grad{-1}_{i-2}}e^{-\psi_{i-2}+2\psi_{i-1}-\psi_{i}}\)\d_{i,k+2} &\\
+{\lambda^{-1}}{\sigma}\({u\grad{-1}_1}e^{-\phi_1-\phi_n} - {u\grad{-1}_{n+1}}e^{2\psi_1-\psi_2}\)\d_{i,2}\d_{k,n+1} &\\
+{\lambda^{-1}}{\sigma}\({u\grad{-1}_{n+1}}e^{-\phi_{n-1}+2\phi_{n}}-{u\grad{-1}_n} e^{-\psi_1-\psi_n}\)\d_{i,1}\d_{k,n} &\\
+\pa_t{u\grad{0}_i} \d_{i,k} &= 0.\\
\end{split}
\label{t01}
\end{equation}

Taking  $i=k-1$  in eqn. (\ref{x01}) and $i=k$ in (\ref{t01}) we find that ${u\grad{0}_1} = {u\grad{0}_2} = \cdots ={u\grad{0}_{n+1}}  = const. \equiv {1}$. For $i=k+1$  in eqn. (\ref{x01}) 
\begin{equation}
{u\grad{-1}_k} ={c_k}\,e^{\phi_k-\phi_{k-1}-\psi_{k+1}+\psi_k}, \quad k=1,\cdots, n
\label{bk}
\end{equation}
where ${c_k}$ are constants and $ \psi_0 = \psi_{n+1} = \phi_{0}= \phi_{n+1} = 0$.  Moreover, for $i=1$ and $k=n+1$  in (\ref{x01}) we find ${u\grad{-1}_{n+1}} = {c_{n+1}} e^{-\phi_n-\psi_{1}}$. The remaining eqns.  in (\ref{t01}), i.e. $i=k+2$,  $i=2, k=n+1$ and $i=1,k=n$ implies that ${c_1} = {c_2} = \cdots = {c_{n+1}} \equiv 1$. The  B\"acklund-gauge transformation is then given by
\begin{equation}
U (\phi_i,\psi_i)= 
\(\begin{array}{ccccc}
{1} & 0 & \cdots& \cdots  &\l^{-1}{\sigma} e^{-\phi_n-\psi_1} \\
{\sigma} e^{\phi_1-\psi_2+\psi_1} & {1} & &  \cdots & 0 \\
0  &  \ddots & \ddots &  \cdots     & \vdots\\
\vdots &  & {\sigma} e^{\phi_k-\phi_{k-1}-\psi_{k+1}+\psi_{k}}. & {1}       &0 \\
0& \cdots & &{\sigma} e^{\phi_n-\phi_{n-1}+\psi_n}  &{1}
\end{array}\).
\label{btg}
\end{equation}
\\
The B\"acklund transformations  are given by setting $i=k$ in  eqn. (\ref{x01}) and $i=k+1$ of eqn. (\ref{t01}), namely
\begin{equation}
\pa_x (\phi_i-\phi_{i-1}-\psi_i+\psi_{i-1}) = {\sigma}( e^{\phi_i-\phi_{i-1}-\psi_{i+1}+\psi_{i}} - e^{\phi_{i-1}-\phi_{i-2}-\psi_i+\psi_{i-1}}), 
\label{btx}
\end{equation}
\begin{equation}
\pa_t (\phi_{i}-\phi_{i-1}-\psi_{i+1}+\psi_{i})={{{1}\over{\sigma}} } (e^{\phi_{i}-\phi_{i+1}+\psi_{i+1}-\psi_{i}} - e^{-\phi_{i}+\phi_{i-1}+\psi_{i}-\psi_{i-1}}),
\label{btt}
\end{equation}
$ i=1,\cdots,n$ and $\phi_{-1}= \phi_n$, together with the border terms
\begin{align}
\pa_x(\phi_n-\psi_n) &= {\sigma}\,( e^{\phi_n-\phi_{n-1}+\psi_n}- e^{-\phi_n-\psi_1}),  \\
\pa_t(\phi_n-\psi_1) &={{{1\over{\sigma}} }}( e^{-\phi_1+\psi_{1}}- e^{\phi_n-\psi_n}) \label{2.18}
\end{align}

It  can be checked directly that the compatibility  of BT eqns. (\ref{btx}) - (\ref{2.18})   leads correctly to  the eqns of motion (\ref{3}). These eqns.  coincide with those  found by Fordy and Gibbons in ref. \cite{fordy}.

\subsection{Type II B\"acklund Transformation}

We now propose an alternative solution for the B\"acklund transformations for the affine Toda models, namely type II  B\"acklund transformation (type II BT). Consider the following affine structure associated to  all grades $0\leq q \leq n+1$ within the matrix $U$ of the form,
\begin{equation}
U = \sum_{q=0}^{n+1} {\sigma^{q}}\, U^{(-q)} =U^{(0)} +{\sigma}\, U^{(-1)} +\cdots+ {\sigma^{n+1}}\, U^{(-n-1)}.
\label{ans}
\end{equation}

According to the grading structure  developed in the Appendix A we start proposing the following ansatz for the matrix representation of the B\"acklund-gauge transformation $U(\phi_i, \psi_i)$ in (\ref{004}),
\begin{equation}
U_{a,b} = (u_b^{(0)} +\l^{-1}\,\,u_b^{(-(n+1))}) \d_{a,b}
+ \sum_{q=1}^{n}\,\,(u_b^{(-q)}\d_{a,b+q} + {\lambda^{-1}u_b^{(-q)}\delta_{a, b-(n+1)+q}})
\label{220}
\end{equation} 
or more explicitly,

\begin{equation*}
U = 
\left( {\begin{array}{cccc}
	u_1^{(0)}                    & 0         & \cdots                     & 0        \\
	{\sigma}u_1^{(-1)}     & u_2^{(0)} & \cdots                     & 0        \\
	\vdots                       &           & \ddots                     & \vdots   \\
	{\sigma^{n}}u_1^{(-n)} & \cdots    & {\sigma}u_{n}^{(-1)} &u_{n+1}^{(0)}
	\end{array} } \right)
+ \l^{-1} \left( {\begin{array}{cccccc}
	{\sigma^{n+1}}u^{(-n-1)}_1 & {\sigma^{n}}u_{2}^{(-n)} & \cdots & {\sigma}u^{(-1)}_{n+1}  \\
	0                     & {\sigma^{n+1}}u_2^{(-n-1)} & \cdots   & {\sigma^{2}}u^{(-2)}_{n+1} \\
	\vdots                & \vdots                           & \ddots   & \vdots                           \\
	0                     &  0                               & 0        & {\sigma^{n+1}}u_{n+1}^{(-n-1)}
	\end{array} } \right).
\label{221}
\end{equation*}

In particular this kind of solution is characterized by the presence of nontrivial diagonal terms  associated to   grade $q=-n-1$, namely $u_a^{(-n-1)}, \; a=1, \cdots, n+1$. Notice that in order to recover the type I BT we  set $u_a^{(-q)}=0$ for $ q >1 $.

From eqn. (\ref{004}) we find for 
\begin{equation}
\begin{split}
[ UA_x (\phi_i)-A_x(\psi_i)U-\pa_xU]_{ik}=&U_{i,k}\pa_x(\phi_k-\phi_{k-1}-\psi_{i}+\psi_{i-1}) \\
&+U_{i,k-1}-U_{i+1,k}+\l(U_{i,n+1}\d_{k,1}-U_{1,k}\d_{i,n+1})-\pa_xU_{i,k}\\
=&0 
\end{split}
\label{xx}
\end{equation}
Similarly,  
\begin{equation}
\begin{split}
[ UA_t (\phi_i)-A_t(\psi_i)U-\pa_tU]_{ik}=& U_{i,k+1} e^{-\phi_{k-1}+2\phi_{k}-\phi_{k+1} }+U_{i-1,k}e^{-\psi_{i-2}+2\psi_{i-1}-\psi_{i}} \\
&+\l^{-1}e^{-\phi_1-\phi_n}U_{i,1}\d_{k,n+1}-\l^{-1}e^{-\psi_1-\psi_n}U_{n+1,k}\d_{i,1}-\pa_t U_{i,k}\\
=&0
\end{split}
\label{tt}
\end{equation}
where $U_{i,k}, i,k = 1, \cdots, n+1$  denote the matrix elements of $U$.   

For $i=k$ in eqn. (\ref{tt}) and taking into account that $U_{k,k+1}$ and $U_{k-1,k}$ are both upper diagonal and henceforth proportional to $\l^{-1}$ we find that the only term independent of $\l$ is  
\br 
\pa_t U_{k,k}^{(0)} = \pa_t u_k^{(0)} = 0, 
\; \quad k=1, \cdots, n+1.
\er
Moreover taking $i=l\;\; k=l+1$ in (\ref{xx}) we find,
\begin{equation}
\begin{split}
&U_{l,l+1}\pa_x (\phi_{l+1}-\phi_l-\psi_l+\psi_{l-1}) +U_{l,l}-U_{l+1,l+1}\\
&+\l( U_{l,n+1}\d_{l+1,1}-U_{{1},l+1}\d_{l,n+1} )-\pa_x U_{l,l+1}=0.
\end{split}
\end{equation}
Again the  only terms independent of $\l$ is $u^{(0)}_{l}-u^{(0)}_{l+1}=0$ and  henceforth $u_{l}^{(0)}= \zeta =$ const., $l=1, \cdots, n+1$. Taking the diagonal terms,  $i=k$ in (\ref{xx}) leads to
\begin{equation}
\begin{split}
&U_{k,k}\pa_x (\phi_{k}-\phi_{k-1}-\psi_k+\psi_{k-1}) +U_{k,k-1}-U_{k+1,k}\\
&+\l ( U_{k,n+1}\d_{k,1}-U_{k,k}\d_{k,n+1} )-\pa_x U_{k,k}=0.
\end{split}
\label{ii}
\end{equation}
Taking the $\l^{-1}$ terms of (\ref{ii})  (lower diag. terms $U_{l+1,l}$ are indep. of $\l$),
\br
u_{k}^{(-n-1)} \pa_x (\phi_{k}-\phi_{k-1}-\psi_k+\psi_{k-1}) - \pa_x u_{k}^{(-n-1)} =0.
\er
Integrating we get 
\br
u_{k}^{(-n-1)} ={c_k} e^{\phi_{k}-\phi_{k-1}-\psi_k+\psi_{k-1}}, \qquad k=1, \cdots , n+1,  \qquad {c_k}= {\rm  const}.
\er
If we  now consider the $k+1,k$ matrix element of (\ref{tt}) 
\br
e^{\phi_{k-1}+2\phi_{k}-\phi_{k+1}}U_{k+1,k+1}- e^{\psi_{k-1}+2\psi_{k}-\psi_{k+1}}U_{k,k} - \pa_t U_{k+1,k} =0
\er
and take  terms  proportional to $\l^{-1}$.   Since $U_{k+1,k}$ is lower diagonal and hence indep. of $\l $,   
\br
e^{\phi_{k-1}+2\phi_{k}-\phi_{k+1}}u^{(-n-1)}_{k+1} - e^{\psi_{k-1}+2\psi_{k}-\psi_{k+1}} u^{(-n-1)}_{k} =0, \qquad k=1, \cdots, n
\er
implies  ${c_1}= {c_2}= \cdots =  {c_{n+1}}\equiv {1} $. The diagonal terms in $U$ are then determined as
\br
U_{k,k}= u_{k}^{(0)} +\l^{-1} {\sigma^{-n-1}} \,u_{k}^{(-n-1)} = {1} +\l^{-1}{\sigma^{-n-1}} e^{\phi_{k}-\phi_{k-1}-\psi_k+\psi_{k-1}},
\label{kk}
\er
for $k=1, \cdots, n+1$. We now proceed in  solving from the second lowest grade  $q=-n$ namely, $U_{a,a+1}$ upwards.  Let  us take the $k,k+1$ matrix elements  of (\ref{xx})
which can be rewritten as
\begin{equation}
\begin{split}
-\pa_x\(U_{k,k+1}e^{-\phi_{k+1}+\phi_k+\psi_k-\psi_{k-1}}\)e^{\phi_{k+1}-\phi_k-\psi_k+\psi_{k-1}}+U_{k,k}-U_{k+1,k+1} \\
+\l(U_{k,n+1}\d_{k+1,1}-U_{1,k+1}\d_{k,n+1})=0.
\end{split}
\end{equation}
For $k\neq 1, n+1$ we find after substituting diagonal elements $U_{a,a}$ in (\ref{kk}),
\br
\pa_x \(U_{k,k+1}e^{-\phi_{k+1}+\phi_k+\psi_k-\psi_{k-1}}\) =\l^{-1}\s \(e^{-\phi_{k-1}+2\phi_k-\phi_{k+1}} - e^{\psi_{k-1}+2\psi_{k}-\psi_{k+1}}\)
\er
which determines   $U_{k,k+1}$ by direct integration of the above  equation.
The next in line to be determined is $U_{k,k+2}$ of grade $q=-n+1$ described recursively from (\ref{xx}) in terms of its previous level, $U_{k,k+1}$ by,
\br
\pa_x \(U_{k,k+2}e^{-\phi_{k+2}+\phi_{k+1}+\psi_k-\psi_{k-1}}\) = U_{k,k+1}-U_{k+1,k+2}     
\er
and so on for $U_{k, k+a}$ in terms of $U_{k,k+a-1}$ and $U_{k+1, k+a+1}$.

\section{The $sl(3)$ Example}

In order  to illustrate  the construction with parametrization \eqref{220}  we now  explicitly develop  the $\lie =sl(3)$ example where $U= U^{(0)}+{\s}U^{(-1)}+{\s^2}U^{(-2)} +{\s^3} U^{(-3)}$, or in matrix form, 
\begin{equation*}
U(\phi_i,\psi_i) =
\left( {\begin{array}{ccc} 
	{1} +\l^{-1} {\s^3} e^{q_1}  &\l^{-1} {\s^2} u_2^{(-2)} & \l^{-1}{\s} u_3^{(-1)} \\[4pt]
	{\s} u_1^{(-1)} & {1}  +\l^{-1} {\s^3} e^{-q_1+q_2 }& \l^{-1}{\s^2} u_3^{(-2)}   \\[4pt]
	{\s^2} u_1^{(-2)}&  {\s} u_2^{(-1)}&    {1}  + \l^{-1} {\s^3} e^{-q_2}\
	\end{array} } \right) 
%\label{33}  this label is aready definde... but this equation seems to have no references to it
\end{equation*}
where  ${q_1}= {\phi_1-\psi_1},\quad  q_2 = \phi_2-\psi_2$. Equations (\ref{xx}) and (\ref{tt}) for $sl(3)$  become:
\begin{eqnarray}
\pa_xq_1 &=& {\s} (-u_3^{(-1)}+u_1^{(-1)}), \nonu \\
\pa_x( q_1- q_2) & = & {\s}(u_1^{(-1)}-u_2^{(-1)}),\label{x1} \\
-\pa_xq_2 & = & {\s} (u_3^{(-1)}-u_2^{(-1)}), \nonu \\[8pt]
\pa_x\(u_3^{(-1)}e^{\phi_2+\psi_1}\) &=& {\s} \(u_3^{(-2)}-u_2^{(-2)}\) e^{\phi_2+\psi_1}\nonu, \\
\pa_x\(u_1^{(-1)}e^{-\phi_1-\psi_1 +\psi_2}\) &=& {\s} \(u_2^{(-2)}-u_3^{(-2)}\)e^{-\phi_1-\psi_1 +\psi_2}  \label{x2}, \\
\pa_x\(u_2^{(-1)}e^{\phi_1-\phi_2 -\psi_2}\) &=& {\s} \(u_1^{(-2)} - u_2^{(-2)} \) e^{\phi_1-\phi_2-\psi_2}, \nonu \\[8pt]
\pa_x\(u_2^{(-2)}e^{\phi_-1-\phi_2+\psi_1}\)&=&\s \( e^{2\phi_1-\phi_2}-e^{2\psi_1-\psi_2} \), \nonu \\
\pa_x\(u_3^{(-2)}e^{\phi_2-\psi_1 +\psi_2}\)&=& \s \( e^{-\phi_1+2\phi_2}- e^{-\psi_1+2\psi_2}\),\label{x3} \\
\pa_x\(u_1^{(-2)}e^{-\phi_1-\psi_2}\)&=& \s \(e^{ -\phi_1-\phi_2}-e^{-\psi_1-\psi_2}\), \nonu
\end{eqnarray}
and
\begin{eqnarray}
\s \pa_tq_1& = &-u_1^{(-2)}e^{-\phi_1-\psi_2}+u_2^{(-2)}e^{\phi_1-\phi_2+\psi_1}, \nonu \\
\s  \pa_t(q_1-q_2)& =& u_2^{(-2)}e^{\phi_1-\phi_2+\psi_1}-u_3^{(-2)}e^{\phi_2-\psi_1+\psi_2}, \label{t1} \\
-\s \pa_tq_2& =& u_1^{(-2)}e^{-\phi_1-\psi_2}-u_3^{(-2)}e^{\phi_2-\psi_1+\psi_2}\nonu \\[8pt]
{\s} \pa_t u_2^{(-2)} &=&u_3^{(-1)}e^{-\phi_1+2\phi_2} - u_2^{(-1)} e^{-\psi_1-\psi_2},\nonu \\
{\s} \pa_t u_3^{(-2)}&=&-u_3^{(-1)}e^{2\psi_1-\psi_2}+u_1^{(-1)}e^{-\phi_1-\phi_2}, \label{t2}  \\
{\s} \pa_tu_1^{(-2)}&=& -u_1^{(-1)} e^{-\psi_1+2\psi_2}+u_2^{(-1)} e^{2\phi_1-\phi_2},\nonu \\[8pt]
{\s} \pa_tu_3^{(-1)}&=& \( e^{-\phi_1-\phi_2}-e^{-\psi_1-\psi_2} \), \nonu \\
{\s} \pa_t u_1^{(-1)}&=&  \( e^{2\phi_1-\phi_2}- e^{2\psi_1-\psi_2}\),\label{t3} \\
{\s} \pa_t u_2^{(-1)}&=&  \(e^{ -\phi_1+2\phi_2}-e^{-\psi_1+2\psi_2} \).  \nonu
\end{eqnarray}
Consistency condition  of  eqns. (\ref{x1})-(\ref{t3}) with  second order eqns. of motion can be verified  by direct calculation.
Integrating eqns. (\ref{x3}) we  obtain  $u_i^{(-2)} $ in terms of two specific solutions $(\phi_j ,\psi_j), j=1,2$ which, when inserted in eqns. (\ref{x2})  gives $u_i^{(-1)}, i=1,2,3$.  It then follows  the BT from eqns. (\ref{x1}).  As for the time  components of (\ref{221}) we start by solving  (\ref{t3}) for $u_i^{(-1)} $ which  leads to  $u_i^{(-2)} $  by solving (\ref{t2}) and then the BT from (\ref{t1}).

Let    us  now consider the following quantities,
\begin{align*}
w_1=& \( u_1^{(-1)} u_2^{(-1)} u_3^{(-1)} - u_1^{(-1)} u_2^{(-2)} - u_2^{(-1)} u_3^{(-2)} - u_3^{(-1)} u_1^{(-2)} + e^{q_1} +e^{-q_2}+e^{-q_1+q_2} \)_ , \\
w_2=& \left( u_1^{(-2)} u_2^{(-2)} u_3^{(-2)} - u_1^{(-1)} u_2^{(-2)} e^{-q_2} + u_3^{(-1)} u_1^{(-2)} e^{-q_1+q_2} + u_2^{(-1)} u_3^{(-2)} e^{q_1}  \right.  \\
& \left.\,\, + \,e^{q_1}+e^{-q_2}+e^{-q_1+q_2} \right) .
\label{w}
\end{align*} 
It can be checked by direct calculation  using (\ref{x1})-(\ref{t3}) that, 
\br \pa_xw_a= \pa_tw_a=0, \quad a=1,2.
\label{www}
\er   
and hence, $ w_1 = \eta_1, \quad w_2  = \eta_2, \quad \eta_a = {\rm const},\; \; a=1,2.$

In particular,  it follows that 
\br
Det [U(\phi_i, \psi_i)]= {1} +\l^{-1}{\s^3} w_1+\l^{-2}{\s^6} w_2 + {\s^9}\l^{-3}. \qquad \label{ww}
\er
At this point we should like to raise the fact   that  fields $u_i^{(-1)}$ and $u_i^{(-2)}, i=1,2,3$  are not all independent.  
In particular for type II BT   the relevant degrees of freedom can be  fixed such that  
\br U(\psi_i,\phi_i) \sim U^{-1}(\phi_i,\psi_i). \label{inv}
\er

We now propose  a  solution  for $U$ incorporating the conservation laws (\ref{ww})  and condition (\ref{inv})  in the lines  suggested in ref. \cite{corrigan}  as
\begin{eqnarray}
U(\phi_i,\psi_i)& = &
\left( {\begin{array}{ccc}
	{1} +\l^{-1}\s^3 e^{q_1}  & \l^{-1}\s^2\,A_1{{e^{-\L_1}}} &   \l^{-1}\s\, C{{e^{-\L_1-\L_2}}} \\[6pt]
	\s\, e^{\L_1}& {1} +\l^{-1}\s^3 e^{-q_1+q_2 } & \l^{-1}\s^2\,A_2e^{-\L_2}  \\[6pt]
	\s^2\,De^{\L_1+\L_2}& \s\, e^{\L_2}&  {1} +\l^{-1}\s^3 e^{-q_2} \\
	\end{array} } \right) 
\label{33}
\end{eqnarray}
where $\L_1$ and $\L_2$ are  two auxiliary fields   and 
\begin{align*}
w_1 &=  \( e^{q_1}+e^{-q_1+q_2}+e^{-q_2} + C- CD - A_1 - A_2 \), \\
w_2 &=  \( e^{-q_1}+e^{q_1-q_2}+e^{q_2} + A_1 A_2 D - e^{-q_2} A_1 - e^{q_1} A_2 - e^{-q_1+q_2} C D \)
\end{align*}  
Here, in order to establish (\ref{www}), (\ref{ww}) it follows from  (\ref{inv}) that, 
\br
A_1   &=& (1+\e_1) (1 +\e_2), \nonu \\
A_2   &=& (1+ \e_2)(1+\e_3), \nonu \\
C     &=& (1+\e_1) (1+ \e_2)(1+\e_3), \nonu \\
D     &=& {{1}\over {(1+\e_2)}},
\label{abcd}
\er
where
\br
\e_1 = e^{q_1}, \qquad \e_2 = e^{-q_1+q_2}, \qquad \e_3 = e^{-q_2}.
\er
Notice that  both $w_1 = w_2  = {-1}$ are independent of  the auxiliary fields $\L_i, \;\; i=1,2$.     
In fact   inverting    matrix (\ref{33})  taking into account (\ref{abcd}) we can verifiy explicitly  that  condition (\ref{inv}) holds with $\s \rightarrow -\s$, i.e.,
\br
U^{-1}(\phi_i,\psi_i,\s )  = {{\l^2}\over {\l^2-\s^2}} U(\psi_i,\phi_i,-\s ).
\label{inv1}
\er

Moreover the B\"acklund  transformation  in terms of variables  defined in (\ref{33}) can be derived from (\ref{x1}) - (\ref{t3}) to be
\begin{eqnarray}
\pa_xq_1 &=&{\s} \( e^{\L_1}-e^{-\L_1-\L_2}C \), \nonu \\
\pa_xq_2 &=& {\s} \( e^{\L_2}-e^{-\L_1-\L_2}C \),  \label{15}\\	
\s \pa_tq_1 &=&  e^{\phi_1-\phi_2+\psi_1}e^{-\L_1}A_1 -   e^{-\phi_1-\psi_2}e^{\L_1+\L_2}D, \nonu \\
\s \pa_tq_2&=& e^{\phi_2-\psi_1+\psi_2}e^{-\L_2}A_2-e^{-\phi_1-\psi_2}e^{\L_1+\L_2}D,  \nonu \\[8pt]
{\s} \pa_t\(e^{-\L_1-\L_2}C\) &=& e^{-\phi_1-\phi_2}-e^{-\psi_1-\psi_2}, \nonu \\
{\s} \pa_t\(e^{\L_1}\)&=& e^{2\phi_1-\phi_2}-e^{2\psi_1-\psi_2}, \label{17} \\
{\s} \pa_t\(e^{\L_2}\)&=& e^{-\phi_1+2\phi_2}-e^{-\psi_1+2\psi_2}, \nonu \\[8pt]
\pa_x\(e^{-\L_1}A_1e^{\phi_1-\phi_2+\psi_1}\) &=&\s (e^{2\phi_1-\phi_2}-e^{2\psi_1-\psi_2}), \nonu \\
\pa_x\(e^{-\L_2}A_2e^{\phi_2-\psi_1+\psi_2}\) &=&\s(e^{-\phi_1+2\phi_2}-e^{-\psi_1+2\psi_2}), \label{3.10} \\
\pa_x\(e^{\L_1+\L_2}De^{-\phi_1-\psi_2}\) &=&\s(e^{-\phi_1-\phi_2}-e^{-\psi_1-\psi_2}).  \nonu
\end{eqnarray}
together with
\begin{eqnarray}
\pa_x\( Ce^{-\L_1-\L_2} e^{\phi_2+\psi_1}\) &=& {\s} e^{\phi_2+\psi_1}\(e^{-\L_1}A_1-e^{-\L_2}A_2\), \nonu \\
\pa_x\(e^{\L_1} e^{-\phi_1-\psi_1+\psi_2}\)  &=& {\s} e^{-\phi_1-\psi_1+\psi_2}\(e^{-\L_2}A_2-e^{\L_1+\L_2}D\), \label{3.11} \\
\pa_x\(e^{\L_2} e^{\phi_1-\phi_2-\psi_2}\)   &=&  {\s} e^{\phi_1-\phi_2-\psi_2}\(-e^{-\L_1}A_1+e^{\L_1+\L_2}D\), \nonu 
\end{eqnarray}
and
\begin{eqnarray}
{\s} \pa_t\(A_1e^{-\L_1} \) &=& e^{-\phi_1+2\phi_2}e^{-\L_1-\L_2}C-e^{-\psi_1-\psi_2}e^{\L_2}, \nonu \\
{\s} \pa_t\(A_2e^{-\L_2} \) &=& e^{-\phi_1-\phi_2}e^{\L_1}-e^{2\psi_1-\psi_2}e^{-\L_1-\L_2}C, \label{3.12} \\
{\s} \pa_t \(De^{\L_1+\L_2}\)&=&e^{2\phi_1-\phi_2}e^{\L_2}-e^{-\psi_1+2\psi_2}e^{\L_1}. \nonu
\end{eqnarray}
It  is straightforward to verify  the consistency  of (\ref{15})-(\ref{3.10})   with eqns. of motion namely,
\br
\pa_t\pa_x \varphi_1 =e^{2\varphi_1-\varphi_2}-e^{-\varphi_1-\varphi_2}, \qquad \pa_t\pa_x \varphi_2 =e^{-\varphi_1+2\varphi_2}-e^{-\varphi_1-\varphi_2}
\label{3.19}
\er
for $\varphi_i = \phi_i$ or $\psi_i$.

Eqns. (\ref{3.11}) and (\ref{3.12})    can be verified  directly  using (\ref{15})-(\ref{3.10}).

\subsection{Composition of B\"acklund Transformation}

From eqn. (\ref{004}) we find that
{
\br
A_{\mu}(\phi_i) =U^{-1}(\phi_i,\psi_i)A_{\mu}(\psi_i)U(\phi_i,\psi_i) +U^{-1}(\phi_i,\psi_i)\pa_{\mu}U(\phi_i,\psi_i) 
\er }
which is equivalent to
\br
A_{\mu}(\psi_i) =U(\phi_i,\psi_i)A_{\mu}(\phi_i)U^{-1}(\phi_i,\psi_i) -\pa_{\mu}U(\phi_i,\psi_i) U^{-1}(\phi_i,\psi_i).
\er
Exchanging $\phi_i \leftrightarrow \psi_i$  we find that
\begin{equation}
\tilde U(\phi_i,\psi_i) = {Det\(U(\phi_i,\psi_i)\)}\,U^{-1} (\psi_i,\phi_i)
\label{tilde}
\end{equation}
also  correspond to a B\"acklund-gauge transformation.  We now {analyse} the various cases of relation \eqref{tilde}.

Consider the type I B\"acklund-gauge transformation  \eqref{btg} which for $\lie =sl(3)$ is given  by
\br
{U^I}(\phi_i,\psi_i, {\s})& = &
\left( {\begin{array}{ccc}
		1  &0 &   \l^{-1} {\s} e^{-\phi_2-\psi_1} \\
		{\s} e^{\phi_1+\psi_1-\psi_2}& 1 & 0\\
		0&   {\s} e^{\phi_2-\phi_1+\psi_2} &1
\end{array} } \right).
\label{33i}
\er
Taking its  inverse, we find from  \eqref{tilde},  that
\begin{equation}
\tilde{U} (\psi_i, \phi_i,\s)\,\, =
\left( {\begin{array}{ccc}
	1  &\l^{-1} {\s^2} e^{-\phi_1-\psi_1+\psi_2} &  - \l^{-1} {\s} e^{-\phi_2-\psi_1} \\
	-{\s} e^{\phi_1+\psi_1-\psi_2}& 1 & \l^{-1}{\s^2}e^{\phi_1-\phi_2-\psi_2}\\
	{\s^2} e^{\phi_2+\psi_1}&   - {\s} e^{-\phi_1+\phi_2+\psi_2} &1
	\end{array} } \right) 
\label{334}
\end{equation}
also satisfy (\ref{004})  and  has  the following grading structure $ \tilde { U}(\phi_i,\psi_i) = U^{(0)}+ \s U^{(-1)} + \s^2 U^{(-2)}$.   
From  such  structure the natural  question whether $\tilde U$ can be  written as a product of two type I BT arises.  In fact  we can verify the  following relation
 {\br
U^I( \theta_i, \phi_i, {\s_1})*U^I(\psi_i, \theta_i, {\s_2}) = \tilde U (\psi_i, \phi_i, {\s} )\label{uu}
\er }
for $ \theta_1 = \phi_2 + \psi_1 - \psi_2, \quad \theta_2 = -\phi_1 + \phi_2 + \psi_1, \quad \s_1 = \omega \s, \quad \s_2 = \omega^{*} \s, \quad \omega^3 = 1$ and $\omega \neq 1$.

Following the same line of thought we ask ourselves whether the  type II gauge-B\"acklund transformation  $U^{II}(\phi_i,\psi_i, \s) $ can be decomposed as product of more fundamental structures.  Indeed, for $sl(3)$  we  find that $U (\phi_i,\psi_i, \s) $ given by eqn. (\ref{33}) can be written as,
\begin{equation}
U^{II} (\phi_i,\psi_i, \s) = \tilde{U}({\theta_i,\psi_i,\s_2}) * U^{I}({\phi_i,\theta_i,\sigma_1})
\end{equation}
and the auxiliary fields $\Lambda_i$ in (\ref{33}) of the type II BT are given by
\begin{align}
\L_1 &= {\theta _1-\theta _2+\ln \left(e^{\psi _1}+e^{\phi _1}\right)}, \label{326}\\
\L_2 &= {\theta _2+\ln \left(e^{\psi _2-\psi _1}+e^{\phi _2-\phi _1}\right)} \label{327}
\end{align}
and ${\sigma_1 = \sigma_2 = \sigma} $.

The above  argument  shows that type II  BT  can  be decomposed  in terms of  product  of 3 type I BT    and the auxiliary  fields  $\L_i, \; i=1,2$  correspond  via (\ref{326}) and (\ref{327}), to intermediate  configurations $\theta_i, \; i=1,2$.   For higher rank algebras  the  number of terms in the product   to construct type II BT increases accordingly (in order to generate  nontrivial  diagonal terms for the type II BT).

\subsection{Tzitz\'eica Limit}

Here we derive the  B\"acklund Transformation for the Tzitz\'eica model proposed in \cite{corrigan} from the $sl(3)$ prototype discussed in the previous section.   Similar approach was employed in terms of {$3\times3 $ }matrices in \cite{thiago}.
Consider the reduction process where  we impose the following constraints
\br
\phi_1=\phi_2\equiv  \phi, \qquad \psi_1=\psi_2\equiv  \psi,
\label{phi}
\er
or equivalently $q_1=q_2\equiv q =\phi -\psi$ and $p = \phi + \psi$.  Eqns. (\ref{3.10}) become,
\br
\pa_xq &=& {\s} \( e^{\L_1}-e^{-\L_1-\L_2}C \)\nonumber \\
       &=& \s \( e^{\L_2}-e^{-\L_1-\L_2}C \), \label{01} \\
\s \pa_tq &=&  e^{\psi}e^{-\L_1}A_1 -   e^{-\phi-\psi}e^{\L_1+\L_2}D\nonumber\\
        &=& e^{\phi}e^{-\L_2}A_2-e^{-\phi-\psi} e^{\L_1+\L_2}D,   \label{02}\\
\s \pa_t\(e^{-\L_1-\L_2}C\) &=& e^{-2\phi}-e^{-2\psi},   \label{03}\\[8pt]
\s \pa_t\(e^{\L_1}\) &=& (e^{\phi}-e^{\psi}) = {\s} \pa_t\(e^{\L_2}\),  \label{04}\\
\pa_x\(e^{-\L_1}A_1e^{\psi}\) &=&\s (e^{\phi}-e^{\psi} )= \pa_x\(e^{-\L_2}A_2e^{\phi}\),  \label{05}\\
\pa_x\( e^{\L_1+\L_2}De^{-\phi-\psi} \) &=& \s (e^{-2\phi}-e^{-2\psi} ), \label{06} \\[8pt]
\s\pa_t\(A_1e^{-\L_1}\) &=& e^{\phi}e^{-\L_1-\L_2}C-e^{-2\psi}e^{\L_2}, \label{07} \\
\s\pa_t\(A_2e^{-\L_2}\) &=& -e^{\psi}e^{-\L_1-\L_2}C+ e^{-2\phi}e^{\L_1}, \label{08} \\
\s\pa_t\(De^{\L_1+\L_2}\) &=& -e^{\psi}e^{\L_1}+e^{\phi}e^{\L_2}.  \label{09}
\er
The second equality in (\ref{01})   as well as   eqn. (\ref{04})   imply $\L_1=\L_2=\L$.
From  (\ref{05}) we  obtain
\br
A_1e^{\psi}= A_2e^{\phi}.
\label{A}
\er
Taking the sum of (\ref{07}) and (\ref{08}) we get
\br
\s \pa_t\(e^{-\L}A_2(e^{\phi-\psi}+1)\) &=&  { C e^{-2 \L} \( e^{\phi} - e^{\psi} \)} +  (e^{-2\phi}  -e^{-2\psi} )e^{\L}, \nonu \\
&=&{\s \, \pa_t (e^{-\L}C)}
\er	
and  obtain
\br
A_2\(1+e^{\phi-\psi}\)=  C.
\er
On other hand, from (\ref{09})  and (\ref{04}),
\br
\pa_t(e^{2\L}D) = e^{\L}(e^{\phi}-e^{\psi}) = {{1\over 2}} \pa_t(e^{2\L})
\er
we therefore find that $D={{1}\over {2}}$.
Under constraints (\ref{phi}) we get from  (\ref{3.11}),
\br
2\pa_x(e^{\L}) - e^{\L}\pa_x(\phi+\psi) = {\s} e^{-\L}(A_2-A_1) = {\s} e^{-\L} A_2(1-e^{\phi-\psi})
\er
which inserted in (\ref{06}) yields,
\br
{{1}\over {2}}\pa_x(e^{2\L}e^{-\phi-\psi})&=& {{1}\over {2}} e^{\L}\( 2\pa_x(e^{\L})-  e^{\L}\pa_x(\phi+\psi)\) e^{-\phi-\psi} \nonu \\
&=& {{{\s}}\over {2}} e^{\L}\( e^{-\L}A_2(1-e^{\phi-\psi})\) e^{-\phi-\psi} \nonu \\
& =&\s  (e^{-2\phi}-e^{-2\psi}),
\er
and 
\br
A_2=2 (1+e^{-q}). 
\er
It then follows that
\br A_1=2 (1+e^{q}) , \qquad C=2 (e^q +e^{-q}+2)
\er
and the  B\"acklund-gauge transformation  for the Tzitz\'eica model is given  by
\br
U(\phi,\psi)& = &
\left( {\begin{array}{ccccc}
		{1} +\l^{-1}{\s^3} e^{q}  && \l^{-1}2 {\s^2} e^{-\L}(1+e^q)&&   \l^{-1}2 \s e^{-2\L}(e^q +e^{-q}+2) \\[5pt]
		{\s} e^{\L}&& 1 +\l^{-1}{\s^3} && \l^{-1}2 {\s^2}e^{-\L} (1+e^{-q}) \\[5pt]
		{{{\s^2}}\over {2 }}e^{2\L}&& {\s} e^{\L}&&  1 +\l^{-1}{\s^3} e^{-q}
\end{array} } \right) 
\label{tzit}
\er
leading to the following first order eqns.
\br
\pa_x q &=&{\s} \( e^{\L} - 2 e^{-2\L}(2+e^q+e^{-q}) \),   \label {tzit1}\\
\s \pa_t q &=& 2 \,{ e^{- \L + {1 \over 2} p}} \(  { e^{- {1 \over 2} q} + e^{+ {1 \over 2} q } } \)- {{1}\over {2}}e^{2\L-p}, \\
{\s} \pa_t \L &=& e^{-\L +{{1}\over {2}}p}(e^{{{1}\over {2}}q} - e^{-{{1}\over {2}}q}), \\
\pa_x (\L-{{1}\over {2}}p) &=& \s \, e^{-2\L}(e^{-q}-e^{q}). 
\label{tzit2}
\er
It follows that eqns. (\ref{tzit1})-(\ref{tzit2})  coincide precisely with those  proposed in ref. \cite{corrigan} when variables  are changed to $ t = -{{1}\over {\sqrt{2}}}x_+ , x = {{1}\over {\sqrt{2}}}x_-, q\rightarrow -2q,  p\rightarrow -2p, e^{\L} \rightarrow 2e^{-\l}$.

Employing the same limiting procedure (\ref{phi})  to obtain  a consistent limit of the type I B\"acklund  transformation for the Tzitz\'eica model   we find $\phi=\psi$ .  This is  expected  and   was  the main motivation   to introduce  the new structure of auxiliary fields in the  type II BT  in ref. \cite{corrigan}.

\section{$A_n$  B\"acklund-gauge Transformation}

Consider the B\"acklund-gauge Transformation $U(\phi_i,\psi_i) $ corresponding to the map $A_{\mu}(\psi_i) \rightarrow A_{\mu}(\phi_i)$
\begin{equation}
A_{\mu}(\phi_i) = U^{-1}(\phi_i,\psi_i) A_{\mu}(\psi_i) U(\phi_i,\psi_i) + U^{-1}(\phi_i,\psi_i) \pa_{\mu} U(\phi_i,\psi_i).
\label{4-001} 
\end{equation}
Conversely, 
\begin{equation}
U(\psi_i,\phi_i) \,\, {\sim} \,\, U^{-1}(\phi_i,\psi_i) 
\label{uu}
\end{equation}
maps   $A_{\mu}(\phi_i) \rightarrow A_{\mu}(\psi_i)$. Since $A_{\mu}$ belongs to a null traced algebra, by Jacobi's formula we find that \eqref{4-001} implies $Det\(U(\phi_i,\psi_i)\) = const. $ and  following our ansatz in (\ref{ans})\\
\begin{equation*}
Det \(U(\phi_i,\psi_i)\) = w_0+ {{{\sigma^{n+1}}\,w_1}\over {\l}} +\cdots + {{{\sigma^{n(n+1)}}\,w_{n}}\over {\l^{n}}} + {{{\sigma^{(n+1)^2}}\,{w_{n+1}}}\over {\l^{n+1}}}  
\label{det}
\end{equation*}
which implies that 
\begin{equation*}
\pa_{x} w_a= \pa_{t} w_a= 0, \qquad a=0,\cdots, n+1.
\end{equation*}
These last eqns indicate that parametrization  \eqref{220}  is such that
\br
w_a[u_j^{(q)}(\phi_i, \psi_i)]= \eta_a= {\rm const.}, \qquad a=0, \cdots, n+1.
\label{eta}
\er

Our conjecture here  states that   for the $\lie= sl(n+1)$ model there should be $n$ auxiliary fields $\L_i, \;i=1, \cdots, n$  and  the B\"acklund-gauge transformation  described by a $(n+1)$-dimensional matrix $U(\phi_i,\psi_i)_{a,b},\;\; a,b=1, \cdots, n+1$ with  entries:

i) Diagonal
\br
U(\phi_i,\psi_i)_{a,a}&=&  \label{uii}1+ {{\s^{n+1}}\over {\l} }\eps_a, \quad \eps_a = e^{q_a -q_{a-1}}, \quad a=1,\cdots, n+1, \quad q_0=q_{n+1}=0.
\er

ii)  Lower diagonal,
\begin{align*}
U(\phi_i,\psi_i)_{a+1,a}&=  \s e^{\L_{a}}, \qquad a=1, \cdots, n; \\
U(\phi_i,\psi_i)_{a+2,a}&= \s^2 {{e^{\L_a+\L_{a+1}}}\over {(1+\eps_{a+1}) }},\qquad a=1, \cdots, n-1\\ % 
\vdots\qquad & \qquad\qquad\vdots \\
U(\phi_i,\psi_i)_{a+l,a}&= \s^{l} {{e^{\L_a \cdots +\L_{a+l}}}\over {(1+\eps_{a+1})\cdots (1+ \eps_{a+l-1} )}},\qquad a=1, \cdots ,n-l+1 \\ 
\vdots \qquad & \qquad\qquad\vdots  \\
U(\phi_i,\psi_i)_{n+1,1}&= \s^{n}{{e^{\L_1 \cdots +\L_{n}}}\over {(1+\eps_{2})\cdots (1+ \eps_{n} )}}. \\ 
\end{align*}

iii) Upper diagonal 
\begin{align*}
U(\phi_i,\psi_i)_{a-1,a}&=\l^{-1} \s^{n} e^{-\L_{a-1}}(1+\eps_{a-1})(1+\eps_{a}), \qquad  a=2, \cdots, n+1 ,\\[4pt]
U(\phi_i,\psi_i)_{a-2,a}&= \l^{-1} \s^{n-1}e^{-\L_{a-2}+\L_{a-1}}(1+\eps_{a-2})(1+\eps_{a-1})(1+\eps_{a}) , \qquad a=3, \cdots, n+1,\\
\vdots\qquad & \qquad\qquad \vdots \nonu \\
U(\phi_i,\psi_i)_{a-l,a}&=\l^{-1}  \s^{n+1-l}e^{-\L_{a-l}\cdots - \L_{a-1}}(1+\eps_{a-l})\cdots (1+\eps_{a}),\qquad a= l+1, \cdots, n+1,\\ 
\vdots\qquad & \qquad \qquad\vdots \nonu \\
U(\phi_i,\psi_i)_{1,n+1}&= \l^{-1} \s e^{-\L_{1}\cdots - \L_{n}}(1+\eps_{1})(1+\eps_{2})\cdots (1+\eps_{n+1}). \\
\end{align*}
This was verified explicitly for $sl(2)$, $sl(3)$ and $sl(4)$. Inverting  $U(\phi_i,\psi_i)$ for these cases    we verify that indeed  $U(\psi_i, \phi_i, -\s)  = c\,U^{-1} (\phi_i,\psi_i,\s)$  where the constant for $\lie =sl(n+1)$ is.
\begin{equation}
c = \(\frac{\lambda}{\lambda-\sigma^{n+1}}\)^{n-1}Det(U\(\phi_i,\psi_i)\).
\end{equation}

\section{Construction of Integrable Hierarchies}

Here we  shall consider generic time evolution equations  which are classified according  to  the grading structure developed in the Appendix A. Consider the decomposition of an affine Lie algebra $\hat \lie = \sum_a \lie_a$ and  a  constant grade one generator $E \equiv E^{(1)}$ which decomposes  $\hat \lie$ into  Kernel  $\hat \ck $  and its complement, $\hat \cm$ i.e.,
\br
\Hat \lie =\hat { \ck} \oplus \hat {\cm}.
\er
In particular, projecting into the zero grade subspace, $\lie_0 = \ck\oplus \cm  $.   Define  now  the Lax operator as
\br
L=\pa_x + E+A_0,
\label{lax}
\er
where $A_0  \in \cm$ (and consequently $A_0 \in \lie_0$).   In the case of principal gradation   with subspaces given in (\ref{a4}), $A_0 $ can be parametrized as $A_0=\sum_{i=1}^n  \pa_x \phi_i h_i^{(0)} \equiv v_i h^{(0)}_i $.
We now propose  the construction of  time evolution equations in the zero curvature representation,
\br
\lb \pa_x+ A_x, \pa_{t_N} + A_{t_N}\rb =\pa_xA_{t_N}- \pa_{t_{N}}A_x  +\lb A_x, A_{t_N} \rb = 0
\label{zcc}
\er
where $A_x = E+ A_0$ and consider two classes of  solutions.  

\begin{itemize}
	
	\item For the {\it negative grade} time evolution  eqns. we  consider the following ansatz,
	\br
	A_{t_{-N} } = D^{(-N)} + D^{(-N+1)} \cdots + D^{(-1)}, \qquad D^{(-j)} \in \lie_{-j}
	\label{aneg}
	\er
	In this case we start solving (\ref{zcc}) from its lowest grade projection, i.e.,
	\br
	\pa_x D^{(-N)} + \lb A_0, D^{(-N)}\rb = 0
	\nonu
	\er
	which yields a nonlocal equation for $D^{(-N)}$.  
	The second lowest projection 
	of grade $-N+1$ leads to 
	\br
	\pa_x D^{(-N+1)} + \lb A_0, D^{(-N+1)} \rb + \lb E^{(1)},D^{(-N)} \rb = 0
	\nonu
	\er
	and determines $D^{(-N+1)}$.  Continuing recursively we end up  with the zero grade component,
	\br
	\pa_{t_{-N}}A_0 - [ E^{(1)}, D^{(-1)}] =0
	\er
	which  yields the equations of motion according to time $t_{-N}$.
	
	A particular interesting  case is for $t_{-N} =t_{-1}$ where $A_{t_{-1} } = D^{(-1)} $.
	The relevant equations to solve are
	\br
	\pa_x D^{(-1)} + \lb A_0, D^{(-1)}\rb & = & 0, \nonu \\ \nonu \\
	\pa_{t_{-1}} A_0  - \lb  E^{(1)},D^{(-1)}\rb  & = & 0.
	\nonu
	\er
    	 {and their  general $t_{-1}$  solution given  in terms of a zero grade group element    $B = \exp (\lie_0)$  by}
	\br
	D^{(-1)} = B^{-1} E^{(-1)} B, \qquad A_0 = B^{-1}\pa_x B. 
		\nonu
	\er
	The associated time evolution (relativistic {\footnote{ If we take $x$ and $t_{-1}$ to be the light cone coordinates}}) is  then given by the Leznov-Saveliev equation (Affine Toda eqns.) \cite{ls}, 
	\br
	\pa_{t_{-1}} \(B^{-1}\pa_x B\) = \lb  E^{(1)}, B^{-1} E^{(-1)}B \rb 
	\nonu
	\er
	for $B=e^{\sum_{=1}^n\phi_a h_a}$ and coincides with eqn. (\ref{2}).
	
	%%%%%%%%%%%%%%%%%%%%%%%%%%%%%%
	\item The {\it  positive grade}  time evolution  eqns. where  $A_{t_N}$ is   given by 
	\br
	A_{t_N} = D^{(N)} + D^{(N-1)} \cdots +D^{(0)}, \qquad D^{(j)} \in \lie_j, \quad N\in Z_{+}.
	\label{atn}
	\er
	The zero curvature representation (\ref{zcc})  decomposes  according  to the graded structure  into
	\br
	[E^{(1)}, D^{(N)} ]&=&0,\nonu \\
	\lb E^{(1)}, D^{(N-1)} \rb + \lb A_0, D^{(N)}\rb +\pa_x D^{(N)} &=&0,\nonu \\
	\vdots  & & \vdots \nonu \\
	\lb A_0, D^{(0)}\rb  +\pa_x D^{(0)} - \pa_{t_N} A_0 &=&0, \label{a5}
	\er 
	and allows  solving for $D^{(j)}$ recursively starting from the highest grade  eqn. in (\ref{a5}).   In particular, the last eqn. in (\ref{a5})  is the only eqn. involving  time 
	derivatives of $A_0$ and can be  regarded as the time evolution for the  fields  parametrizing  $\cm$.  
	
	The relevant algebraic structure  classifying  the construction of integrable hierarchies are therefore  characterized  by the affine algebra, $\hat \lie$, the grading operator $Q$ and the constant grade one generator $E^{(1)}$. Following the same line of thought developed for $\lie = sl(2)$ we shall consider explicitly  the $\lie= sl(3)$ case as a prototype and consider as an example the first nontrivial positive grade time evolution for $N=2$.
	
	Let $A_0 =v_1h_1^{(0)} +  v_2h_2^{(0)}, \quad    A_{t_2}= D^{(2)} + D^{(1)} + D^{(0)}$ and  solve recursively  the following eqns,
	\br
	[E^{(1)}, D^{(2)} ]&=&0,\nonu \\
	\lb E^{(1)}, D^{(1)} \rb + \lb A_0, D^{(2)}\rb +\pa_x D^{(2)} &=&0, \nonu \\
	\lb E^{(1)}, D^{(0)} \rb + \lb A_0, D^{(1)}\rb +\pa_x D^{(1)} &=&0,\label{a22}\\
	\lb A_0, D^{(0)}\rb  +\pa_x D^{(0)} - \pa_{t_2} A_0 &=&0. \nonu
	\er
	Solving (\ref{a22}) according to the grading structure given in (\ref{a4}) we find
	\br
	D^{(2)} &=&d(E_{-\a_1}^{(1)}+E_{-\a_2}^{(1)}+E_{\a_1+\a_2}^{(0)} ), \qquad d= {\rm const.}\nonu \\
	D^{(1)} &=&  d\(v_2E_{\a_1}^{(0)}   - v_1E_{\a_2}^{(0)} +(v_1-v_2)E_{-\a_1-\a_2}^{(1)}   \) \nonu \\
	D^{(0)} &=&{{1}\over {3}}d \(-\pa_xv_1+2\pa_xv_2+v_1^2-2v_2^2+2v_1v_2\)h_1^{(0)} \label{5.8} \\
	&+&{{1}\over {3}}d\(-2\pa_xv_1+\pa_xv_2+2v_1^2-v_2^2-2v_1v_2\)h_2^{(0)}. \nonu
	\er 
	It then  follows the  time evolution eqns,
	\begin{equation}
	\begin{split}
	\pa_{t_2} v_1 =& {{1}\over {3}}d\pa_x\(-\pa_xv_1+2\pa_xv_2+v_1^2 - 2v_2^2 + 2v_1v_2\),  \\
	\pa_{t_2} v_2 =& {{1}\over {3}}d\pa_x\(-2\pa_xv_1+\pa_x v_2+2v_1^2 - v_2^2 - 2v_1v_2\), \\
	\label{eqmot}
	\end{split}
	\end{equation}
	where $d$ is a constant which from now on  we set $d=1$.
\end{itemize}
 
If we now employ the  type II  B\"acklund-gauge transformation $U(\phi_i, \psi_i, \s)$   proposed in  (\ref{33})  to   transform  the gauge 
potential $A_{t_2}$  with  graded components  given in (\ref{5.8}), which in matrix form   reads,
\br
A_{t_{2}} & = &
\left( {\begin{array}{ccc}
		\pa_{t_{2}}\phi_1 &v_2 &   1 \\
		\l& \pa_{t_{2}}(-\phi_1+\phi_2) & -v_1\\
		\l(v_1-v_2)&  \l&-\pa_{t_{2}}\phi_2
\end{array} } \right) 
\label{33ii}
\er
where $v_1=\pa_x\phi_1, \; v_2=\pa_x\phi_2, \; u_1 =\pa_x\psi_1, \; u_2 = \pa_x \psi_2$.  Acting with (\ref{33}) in eqn (\ref{004}) for $A_{\mu}= A_{t_2}$  in (\ref{33ii})  in terms of matrix elements $(ij)$, we find the following equations,
\br
(11):   \pa_{t_2}q_1 &=& \s ( (-u_1+u_2)Ce^{-\L_1-\L_2}+v_2e^{\L_1}) +\s^2 (De^{\L_1+\L_2}-A_1e^{-\L_1}) \nonu \\
(33): \pa_{t_2}q_2 &=& -\s((v_1-v_2)Ce^{-\L_1-\L_2}+u_1 e^{\L_2})-\s^2(A_2e^{-\L_2}-De^{\L_1+\L_2}) \nonu \\
(21):  \pa_{t_2}\L_1 &=  & \pa_{t_2}(\phi_1+\psi_1-\psi_2) +\s\((u_1-u_2)A_2e^{-\L_1-\L_2}+v_1De^{\L_2}\)\nonu \\
&+&\s^2 e^{-\L_1}(e^{-q_1+q_2}- e^{q_1}) \nonu \\
(32): \pa_{t_2}\L_2 &=& \pa_{t_2}(-\phi_1+\phi_2+\psi_2) +\s\((v_2-v_1)A_1e^{-\L_1-\L_2}+u_2De^{\L_1}\)\nonu \\
 &+&\s^2e^{-\L_2}(e^{-q_2}-e^{-q_1+q_2}), \nonu \\
\label{5.11}
\er
together with
\br
(22):&   \pa_{t_2}(q_1 -q_2)&=\s(u_2e^{\L_1}+v_1e^{\L_2})- \s^2(A_1 e^{-\L_1} -A_2 e^{-\L_2}), \nonu \\
(12) \l^{-1}: &  \pa_{t_2}A_1-A_1\pa_{t_2}\L_1&=A_1\pa_{t_2}(-\phi_1+\phi_2-\psi_1) + \s e^{\L_1}(u_2e^{q_1}-v_2 e^{-q_1+q_2}), \nonu \\
\l^{0}: &  u_2-v_2 &=\s(e^{\L_2}-Ce^{-\L_1-\L_2}), \label{5.12} \\
(23)\l^{-1}: &\pa_{t_2}A_2 -A_2 \pa_{t_2}\L_2 &= A_2 \pa_{t_2}(-\phi_2+\psi_1-\psi_2)+\s e^{\L_2}(v_1e^{-q_2}-u_1e^{-q_1+q_2}), \nonu \\
\l^{0}:&   u_1-v_1 &=\s (e^{\L_1}-Ce^{-\L_1-\L_2}), \nonu 
\er
and 
\br
(31)\l^{-1}: \pa_{t_2}D+D\pa_{t_2}(\L_1+\L_2)&=&A_2\pa_{t_2}(-\phi_2+\psi_1-\psi_2)\nonu \\
&+&\s e^{-\L_1-\L_2}\((u_1-u_2)e^{-q_2}+(v_1-v_2)e^{q_1}\), \nonu \\
\l^{0}:  u_1-u_2-v_1-v_2 &=&\s(e^{\L_1}-e^{\L_2}), \nonu \\
(13)\l^{0}: \pa_{t_2}C-C\pa_{t_2} (\L_1+\L_2)&=&C\pa_{t_2}(-\phi_2-\psi_1)-\s(u_1A_1e^{\L_2}+v_2A_2e^{\L_1})\nonu \\
&+&\s^2 e^{\L_1}(e^{q_1}-e^{-q_2}) \nonu \\
\label{5.13}
\er
Using eqns. (\ref{15}) and (\ref{5.11}) we  can  show that  all eqns. (\ref{5.12}) and (\ref{5.13}) are   identically satisfied.

Also from  (\ref{15}) and (\ref{5.11}), the   equations of motion  (\ref{eqmot}) for fields $v_i$ and $u_i, i=1,2$ can be   verified (see  Appendix B).

\section{Solutions}
%%%%%%%%%%%%%%%%%%%%%%%%%%%%%%%%%%%%
 {In this section we  will discuss  some simple solutions of B\"acklund transformation given by (\ref{15})-(\ref{3.12}). 
    They  are specified in terms of a particular pair of  field configurations,  $(\phi_1, \phi_2)$ and $(\psi_1, \psi_2)$,  both satisfying the equations of motion (\ref{3.19}) and  connected by the BT  (\ref{15})-(\ref{3.12}).  They  are  constructed as follows.}
Consider the first 4 eqns. in \eqref{15} which can be rearranged in a form of  two  systems  of algebraic eqns. for variables $ e^{\L_1} $ and   $ e^{\L_2} $ namely,

\begin{eqnarray}
	a_1e^{\L_1}+b_1&=c_1 e^{-\L_2}, \qquad   a_2e^{\L_1}+b_2=c_2 e^{-\L_2}, \label{sist1}. \\[8pt]
	a_3e^{\L_2}+b_3&=c_3 e^{-\L_1}, \qquad   a_4e^{\L_2}+b_4=c_4 e^{-\L_1}.  \label{sist2}
\end{eqnarray}
where,
\begin{align*}
	&a_1= \frac{1}{\sigma}\partial_x q_2D\,e^{-\phi_1-\psi_2},&  &b_1 = \sigma \pa_tq_2,&  &c_1= A_2 e^{\phi_2-\psi_1+\psi_2}-C D\,e^{-\phi_1-\psi_2},&\\
	&a_2= 1 - \frac{CD}{A_1}e^{-2\phi_1 + \phi_2 -\psi_1-\psi_2},& \qquad &b_2 = -\frac{1}{\sigma}\pa_xq_1,& \qquad &c_2=\sigma\partial_tq_1\frac{C}{A_1}e^{-\phi_1+\phi_2-\psi_1},&\\
	&a_3=\frac{1}{\sigma}\partial_xq_1D\,e^{-\phi_1-\psi_2},& &b_3 =\sigma\pa_tq_1,& &c_3= A_1e^{\phi_1-\phi_2+\psi_1}-CD\,e^{-\phi_1-\psi_2},& \\
	&a_4 =1-\frac{CD}{A_2}e^{-\phi_1-\phi_2+\psi_1-2\psi_2},& &b_4 =-\frac{1}{\sigma}\pa_xq_2,& &c_4 =\sigma\partial_tq_2\frac{C}{A_2}e^{-\phi_2+\psi_1-\psi_2}.&
\end{align*}

The two eqns. in \eqref{sist1} is a system of  equations for the two variables, $X_1 \equiv e^{\L_1}$ and $Y_1 \equiv e^{-\L_2}$ which have solution  given by,
\begin{equation*}
	X_1 ={{b_1c_2-b_2c_1}\over {a_2c_1-a_1c_2}}, \qquad   Y_1={{a_1b_2-a_2b_1}\over {a_1c_2-a_2c_1}}.
\end{equation*}
Likewise \eqref{sist2} is another  system for  variables $X_2 \equiv e^{-\L_1}$  and $Y_2 \equiv e^{\L_2}$ leading to
\begin{equation*}
	X_2 ={{a_3b_4-a_4b_3}\over {a_3c_4-a_4c_3}}, \qquad  Y_2  ={{b_3c_4-b_4c_3}\over {a_4c_3-a_3c_4}}.
\end{equation*}
Consistency  of these  four expressions is given by the compatibility relations,
\begin{equation}
	X_1 X_2 =  e^{\L_1} e^{-\L_1} =1 \qquad   {\rm and}     \qquad  Y_1 Y_2 =   e^{\L_2} e^{-\L_2} =1.
	\label{sec-6-001}
\end{equation}
%where,  after inserting soliton solutions and after some algebraic manipulations, lead us to relations between the soliton momentum $k$, the B\"acklund parameter $\sigma$ and the phase shift $R$, in case of one-one soliton scattering.

%%%%%%%%%%%%%%%%%%%%%%%%%%%%%%%%%%%%%%%%%%%%%%%%%%%%%%%%%%
\subsection{Vacuum-one Soliton solution }
The first   example to be considered is the {\it vacuum $\rightarrow$  one soliton } solution where we set the fields to be
\begin{equation}
	\phi_i=0, \qquad \psi_1 =\ln\({{1+\rho}\over {1+\omega\rho}}\), \qquad   \psi_2 =\ln\({{1+\rho}\over {1+\omega^2\rho}}\).
\end{equation}
    This soliton solutions satisfies the equation (\ref{3}) { with } $\rho=e^{\eta (kx+k^{-1}t)}$,  $\eta^2 =3$ and $\omega^3=1$ \quad $(\omega\neq1)$. The compatibility \eqref{sec-6-001} show that B\"acklund parameter and the momentum are not all independent, instead they satisfy
\begin{equation}
	k^6+\sigma^6  = 0.
\end{equation}

%%%%%%%%%%%%%%%%%%%%%%%%%%%%%%%%%%%%%%%%%%%%%%%%%
\subsection{Scattering of one-Soliton solutions}
%%%%%%%%%%%%%%%%%%%%%%%%%%%%%%%%%%%%%%%%%%%%%%%%%
 {Here we  will consider  the  case in which the two  pairs,  $(\phi_1, \phi_2)$ and $(\psi_1, \psi_2)$ correspond to one soliton solutions.   
    They differ by a phase shift described by the multiplication factor $R$ and can be interpreted as  the scattering of one soliton solutions by the defect.}

The  {\it one soliton  $\rightarrow$  one soliton }  case is a more interesting  since  it is possible to find the phase shift for the soliton scattering.  Let us define
\begin{equation*}
	\phi_1 = \ln\({{1+R\rho}\over {1+\omega R\rho}}\), \quad   \phi_2 =\ln\({{1+R\rho}\over {1+\omega^2R\rho}}\), \qquad
	\psi_1 = \ln\({{1+\rho}\over {1+\omega\rho}}\), \quad   \psi_2 =\ln\({{1+\rho}\over {1+\omega^2\rho}}\) 
\end{equation*}
where $R$ is some complex number namely, the phase shift. From the compatibility conditions \eqref{sec-6-001} it follows that
\begin{equation*}
	\eta  k^6 (R-1)^3+36 k^3 R (R+1) \sigma ^3+\eta  (R-1)^3 \sigma ^6 =0
\end{equation*} 
which admits  3 solutions: 
\begin{align*}
	R_1&=\frac{\alpha - 6\, \beta \,\gamma^{-1/3} + 6\,\gamma^{1/3}}{\eta(k^6 + \sigma^6)}, \\[6pt]
	R_2&=\frac{\alpha - 6\, \omega^2 \beta \gamma^{-1/3} + 6\,\omega \gamma^{1/3}}{\eta(k^6 + \sigma^6)}, \\[6pt]
	R_3&=\frac{\alpha - 6\, \omega  \beta \gamma^{-1/3} + 6\, \omega^2\gamma^{1/3}}{\eta(k^6 + \sigma^6)}, 
\end{align*}
where 
\begin{align*}
	\alpha &= \eta  \sigma ^6+\eta  k^6-12 k^3 \sigma ^3, \\[6pt]
	\beta  &= \eta  k^9 \sigma ^3-4 k^6 \sigma ^6+\eta  k^3 \sigma ^9, \\[6pt]
	\gamma &= \left(\sqrt{k^6 \sigma ^6 \left(k^{12}-\sigma ^{12}\right)^2}-k^3 \sigma ^3 \left(k^{12}-6 \eta  k^9 \sigma ^3+18k^6 \sigma ^6-6 \eta  k^3 \sigma ^9+\sigma ^{12}\right)\right)/2 .
\end{align*}
It can be verified that $R_1\cdot R_2\cdot R_3=1$.

\section{Conclusion and Further Developments}
%%%%%%%%%%%%%%%%%%%%%%%%%%%%%%%%%%%%%%%%%%%%
In this paper we have proposed a systematic construction  of B\"acklund transformation from  gauge transformation acting upon the Lax  pair  and zero curvature representation.  
Such construction  classifies  the various  types of BT in terms  of an affine graded structure.  It also  induces  how  these  constructions can be composed  to produce  more complex  structures.  

From explicit examples of $\lie = sl(3)$ and $sl(4)$,  we observed  the  unexpected feature that $U(\phi_i, \psi_i,\l)=U^{-1}(\psi_i, \phi_i,\l)$,  which    guided us to propose a more general  ansatz for  type II BT for $sl(n+1)$. 
In particular the type II BT for the Tzitz\'eica  model proposed in \cite{corrigan} was  obtained  as a  limiting case of  the BT   for $\lie = sl(3)$.   The Tzitz\'eica  model is an example  where  non-trivial type I BT   does not exist and   
most probably, is a consequence of the twisted   underlying affine  structure.    
Under  this point of view, Toda models associated to algebras  other than $A_n$    has been  recently studied   in  \cite{robertson}, \cite{bristow} and it would be interesting  to  see how  the B\"acklund-gauge construction  could be employed.

Another important aspect of our approach   shows that  the B\"acklund-gauge transformation method   is  universal  in the sense  that it extends  to all equations  within  the hierarchy.  
Since the Lax  operator ($L= \pa_x + E^{(1)} + A_0$) is  the same to all flows  (time evolutions) the $x-$ component of the BT  is   common to all  equations of motion.
A systematic derivation of BT  for  higher flows  can be obtained  directly by gauge transformation  of the  time component $A_{t_{N}}$.

Finally the same method  can be extended to other  integrable hierarchies  as multicomponent AKNS,  Yajima-Oikawa, etc associated to  homogeneous and  mixed  gradation respectively \cite{aratyn-nls},\cite{yo}.

%%%%%%%%%
\vspace{5mm}
{\bf Acknowledgments}
JFG and AHZ thank CNPq and FAPESP for partial  support. JMCF and GVL thank CNPq and Capes for  financial support respectively. 
\appendix

\section{ Affine Algebraic Structure}

Here we shall discuss  the structure of affine Lie algebras and the construction and classification of  integrable hierarchies.   
Consider  an  affine Kac-Moody algebra   $\hat \lie$ defined by 
\br
\lb H_i^{(m)}, H_j^{(n)}\rb &=& \kappa m \d_{i,j} \d_{m+n,0}\nonu \\
\lb H_i^{(m)}, E_{\a}^{(n)}\rb &=&\a^{i}E_{\a}^{(m+n)}\nonu \\
\lb E_{\b}^{(m)}, E_{\a}^{(n)}\rb& =&
\begin{cases}
	\eps (\a, \b)E_{\a+\b}^{(m+n)}, \quad \a+\b = {\rm root},\nonu \\
	{{2}\over {\a^2}}\a \cdot H^{(m+n)} + \kappa m  \d_{m+n,0}, \quad \a+\b =0,\nonu \\
	0\quad  {\rm otherwise}.
\end{cases}
\label{a1}
\er
Define the grading operator $Q$ that decomposes the  affine algebra $\hat \lie$ into grades subspaces, $\lie_a$,
\br 
\hat \lie = \sum_{a\in Z} \lie_a, \qquad \lb Q, \lie_a\rb = a \lie_a, \qquad \lb \lie_a, \lie_b\rb \in \lie_{a+b}
\label{a2}
\er
In this paper we  discuss the $\hat \lie = \hat {sl}(n+1)$ and employ the  principal gradation in which
\br Q = (n+1)d + \sum_{a=1}^{n} \mu_a\cdot H
\label{a3}
\er 
where  $d$ is the derivation operator, i.e.,
\br
\lb d, T_i^{(m)}\rb = m  T_i^{(m)}, \qquad   T_i^{(m)} =  h_i^{(m)} \; {\rm or }\;\; E_{\a}^{(m)}
\er
and
\br
\lb \mu_a \cdot H^{(m)}, E_{\a}\rb = ({\mu_a \cdot \a }) E_{\a}.
\er
Here $\mu_a$  and $\a_a$ are  the fundamental weights and  simple roots respectively,
${\mu_a \cdot \a_b} = \d_{a,b}$, \quad $a,b=1,\cdots, n$, and  have normalized all roots  of $\hat sl(n+1)$ such that $\a^2=2$. 
The  operator $Q$ in (\ref{a3})  induces the following graded subspaces,
\br
\lie_{m(n+1)} &=&\{ h_1^{(m)}, \cdots ,  h_n^{(m)}\},\nonu \\
\lie_{m(n+1) +1} &=&\{ E_{\a_1}^{(m)}, \cdots ,  E_{\a{_n}}^{(m)}, E_{-\a_1\cdots -\a_n}^{(m+1)}\},\nonu \\
\lie_{m(n+1)+2} &=&\{ E_{\a_1+\a_2}^{(m)}, E_{\a_2+\a_3}^{(m)}, \cdots ,  E_{\a_{n-1}+\a{_n}}^{(m)}, E_{-\a_1\cdots -\a_{n-1}}^{(m+1)},   E_{-\a_2\cdots -\a_{n}}^{(m+1)}\},\label{a4} \\
\vdots &=& \vdots \nonu \\
\lie_{m(n+1) +n }&=&\{ E_{-\a_1}^{(m+1)}, \cdots ,  E_{-\a{_n}}^{(m+1)}, E_{\a_1 +\cdots+ \a_n}^{(m)}\}. \nonu
\er
where $h_i^{(m)} = \a_i\cdot H^{(m)}$.

\section{Appendix B}

Here we show the compatibility  of the B\"acklund transformation  for  $t=t_2$. given by eqns. (\ref{15} ), (\ref{3.10})  and (\ref{5.11}) .    From   the equations of motion  (\ref{eqmot})  we find
\br
\pa_{t_2}(u_1-v_1)&=&  
{{1}\over {3}}d\pa_x [ -\pa_x (u_1-v_1)+2\pa_x (u_2-v_2) \nonu \\
&+&(u_1-v_1)(u_1+v_1+2v_2)-2(u_2-v_2)(u_2+v_2-u_1)] \label{xxxx}
\er
From the $x$ component of BT (\ref{15}),
\begin{equation}
\begin{split}
\pa_xq_1 &= -(u_1-v_1) ={\s} \( e^{\L_1}-e^{-\L_1-\L_2}C \), \\
\pa_xq_2 &=-(u_2-v_2)= {\s} \( e^{\L_2}-e^{-\L_1-\L_2}C \),  	
\end{split}
\label{15p}
\end{equation}
and  (\ref{3.10})
\begin{eqnarray}
	\pa_x\( Ce^{-\L_1-\L_2} e^{\phi_2+\psi_1}\) &=& {\s} e^{\phi_2+\psi_1}\(e^{-\L_1}A_1-e^{-\L_2}A_2\), \nonu \\
	\pa_x\(e^{\L_1} e^{-\phi_1-\psi_1+\psi_2}\)  &=& {\s} e^{-\phi_1-\psi_1+\psi_2}\(e^{-\L_2}A_2-e^{\L_1+\L_2}D\), \label{3.11p} \\
	\pa_x\(e^{\L_2} e^{\phi_1-\phi_2-\psi_2}\)   &=&  {\s} e^{\phi_1-\phi_2-\psi_2}\(-e^{-\L_1}A_1+e^{\L_1+\L_2}D\), \nonu 
\end{eqnarray}
we can evaluate 
\br
\pa_x(u_1-v_1)&= &-\s^2 \(2A_2e^{-\L_2}-De^{\L_1+\L_2}-A_1e^{-\L_1} \) \nonu \\
&+&\s e^{\L_1}(-u_1-v_1+v_2)-\s Ce^{-\L_1-\L_2}(u_2+v_1)\nonu \\
\pa_x(u_2-v_2)&=& \s^2 \(-2A_1e^{-\L_1}+De^{\L_1+\L_2}+A_2e^{-\L_1} \) \nonu \\
&+&\s e^{\L_2}(u_1-u_2-v_2)-\s Ce^{-\L_1-\L_2}(u_2+v_1)\nonu \\
\label{pax}
\er
Inserting (\ref{15p}) and (\ref{pax}) in (\ref{xxxx}) we  get the $t_2$  component of BT (first eqn. in  (\ref{5.11})), 
\br
\pa_{t_2}q_1 = \s ( (-u_1+u_2)Ce^{-\L_1-\L_2}+v_2e^{\L_1}) +\s^2 (De^{\L_1+\L_2}-A_1e^{-\L_1}) \nonu 
\er
 Similarly for $\pa_{t_{2}}q_2$ in second eqn. in (\ref{5.11}).

%\acknowledgments

%\paragraph{Note added.} This is also a good position for notes added
%after the paper has been written.

\end{document}